\documentclass{agujournal}

\journalname{JGR-Space Physics}

\usepackage{url}

\begin{document}

\title{
An empirical modification of the force field approach to describe the modulation of galactic cosmic rays close to Earth in a broad range of rigidities
}

\authors{J.~Gieseler\affil{1}, B.~Heber\affil{1}, and K.~Herbst\affil{1} }

\affiliation{1}{Institute of Experimental and Applied Physics, University of Kiel, Leibnizstr. 11, 24118 Kiel, Germany}

\correspondingauthor{J.~Gieseler}{gieseler@physik.uni-kiel.de}

\begin{keypoints}
\item Demonstrate significant rigidity dependence of force field approach
\item Introduce two parameter modification as simple and sufficient workaround
\item Provide rigidity-dependent solar modulation potential for 1973-2016
\end{keypoints}

\begin{abstract}
On their way through the heliosphere, Galactic Cosmic Rays (GCRs) are modulated by various effects before they can be detected at Earth. This process can be described by the Parker equation, which calculates the phase space distribution of GCRs depending on the main modulation processes: convection, drifts, diffusion and adiabatic energy changes. A first order approximation of this equation is the force field approach, reducing it to a one-parameter dependency, the solar modulation potential $\phi$. Utilizing this approach, it is possible to reconstruct $\phi$ from ground based and spacecraft measurements. However, it has been shown previously that $\phi$ depends not only on the Local Interstellar Spectrum (LIS) but also on the energy range of interest. We have investigated this energy dependence further, using published proton intensity spectra obtained by PAMELA as well as heavier nuclei measurements from IMP-8 and ACE/CRIS. Our results show severe limitations at lower energies including a strong dependence on the solar magnetic epoch. Based on these findings, we will outline a new tool to describe GCR proton spectra in the energy range from a few hundred MeV to tens of GeV over the last solar cycles. In order to show the importance of our modification, we calculate the global production rates of the cosmogenic radionuclide $^{10}$Be which is a proxy for the solar activity ranging back thousands of years.
\end{abstract}

\section{Introduction}
\label{section:introduction}
During the last years major progress has been achieved concerning the modulation of Galactic Cosmic Rays (GCRs) due to several facts: 
\begin{enumerate} 
\item Voyager 1 and 2 passed the termination shock at 94 AU \citep{Stone-etal-2005} and 84 AU \citep{Richardson-etal-2008}, respectively, and Voyager 1 the heliopause at 121 AU \citep{Gurnett-etal-2013}, setting the boundary of the modulation volume that is directly influenced by the Sun's activity;
\item the Local Interstellar Spectra (LIS) of ions and electrons are now much better known than ever before \citep{Potgieter-etal-2015,Vos-Potgieter-2015,Bishoff-etal-2016,Corti-etal-2016,Ghelfi-etal-2016, Herbst-etal-2017}	thanks to the Voyager measurements \replaced{[Stone et al., 2013]}{\citep{Stone-etal-2013,Cummings-etal-2016}} in the outer heliosheath and the precise measurements by the PAMELA \citep{Adriani-etal-2011a,Adriani-etal-2011b} and AMS-02 \citep{Aguliar-etal-2015} investigations;
\item the advanced understanding of particle wave interaction in the solar wind that leads to particle scattering, described in the transport equation of \citet{Parker-1965} by diffusive processes \citep{Burger-etal-2000,Tautz-etal-2014,Shalchi-2015}; and
\item the progress in modeling the background heliosphere \citep[e.g.][]{Scherer-etal-2011} and the particle propagation \citep{Potgieter-2013} thanks to increasing computing power. 
\end{enumerate} 

\begin{figure}
\includegraphics[viewport=12 0 766 557, clip, width=\columnwidth]{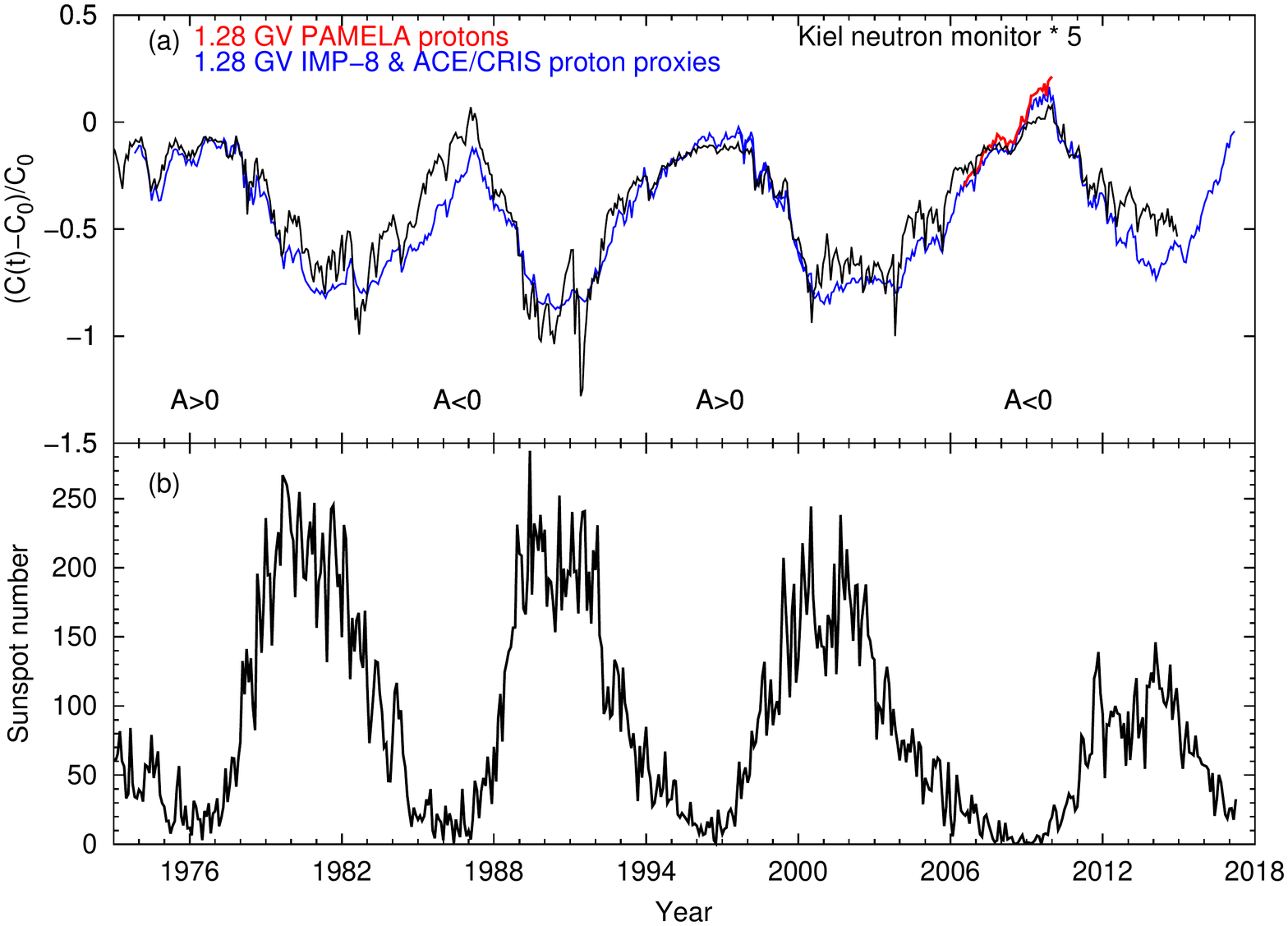}
\caption{Top: Monthly averaged count rate variation of the Kiel neutron monitor (black curve, multiplied by 5 to match scale), intensity variations of 1.28~GV proton measurements by PAMELA (red curve), and the intensity variations of 1.28~GV proton proxies (blue curves, cf. Sect.~\ref{section:proton-proxies}). The data have been normalized to January 2009. Bottom: Monthly sunspot number from the Royal Observatory of Belgium.
}
\label{fig:jgr_icrc2015_sunspot}
\end{figure}

However, another line of research contributed much in our current understanding of GCRs. Cosmogenic radionuclides are the only window to the Sun's activity history over more than a few thousand years. Thus, tremendous effort has been undertaken in order to analyze the different data sets in order to determine the modulation parameter during the Holocene \citep{Vonmoos-etal-2006, Steinhilber-etal-2008, Steinhilber-etal-2012, Herbst-etal-2010}. 
For such studies the first-order force field approximation depending only on one parameter, the force field parameter or solar modulation potential $\phi$, is utilized in order to describe the energy spectra at Earth. 
Commonly, these $\phi$-values are determined using the count rates of neutron monitors \citep{Usoskin-etal-2005,Usoskin-etal-2011,Ghelfi-etal-2017,Usoskin-etal-2017}.
The energy dependent response of such ground based stations at sea level to protons and $\alpha$-particles has recently be investigated by \citet{Mishev-etal-2013}. 
Their Fig.~3 shows that the response is decreasing with decreasing energy, with significantly small contributions below a few GeV.
Taking into account that the GCR spectra of protons and $\alpha$-particles are not strongly energy-dependent in this energy range, neutron monitors are marginally sensitive to energies below a few GeV. 
However, note that during a Ground Level Enhancement (GLE) the energy spectrum of solar energetic particles in the range above 700~MeV is $\propto E^{-3}$ \citep{Mewaldt-etal-2012,Kuehl-etal-2016b}, leading to the fact that in these cases an enhancement is usually measurable \citep{Thakur-etal-2016}. 
Like most spacecraft measurements that are only sensitive to energies below a few GeV the production of cosmogenic radionuclides is sensitive to particles with smaller energies \added{\citep{Webber-Higbie-2003,Webber-Higbie-2010}}. Figure~\ref{fig:jgr_icrc2015_sunspot} (top) shows the count rate variation of the Kiel neutron monitor (black curve, multiplied by 5 to match scale), and the intensity variations of 1.28~GV proton measurements by PAMELA (red curve) as well as 1.28~GV proton proxies (blue curve). All variations have been normalized to January 2009. The bottom panel displays the sunspot number from the Royal Observatory of Belgium. 
The solar magnetic epoch is indicated by A$<$0 and A$>$0, respectively. In an A$>$0-solar magnetic epoch the magnetic field is pointing outward over the northern and inward over the southern hemisphere; vice versa for an A$<$0-solar magnetic epoch. 
Note that here and in the following we move from the energy to the rigidity frame in order to compare measurements of different particle species. A more detailed description is given in Sect.~\ref{section:proton-proxies} where the derivation of the 1.28~GV proton proxies is described.
From Fig.~\ref{fig:jgr_icrc2015_sunspot} it is evident that:
\begin{enumerate}
\item the GCR intensity is anti-proportional to the sunspot number (i. e. the intensity is high when the sunspot number is low and vice versa),
\item the amplitude of the variation is much larger for the lower rigidities than for the higher rigidities (about a factor of 5 when comparing 1.28~GV protons and their proxies with the Kiel neutron monitor), and
\item there is an rigidity dependent difference for an A$>$0 and A$<$0-solar magnetic epoch minimum, i.~e. the intensities are larger for the Kiel neutron monitor in 1987 compared to 1976 and 1997 and vice versa for the 1.28~GV proton proxies (omitting the unusual minimum 2009).
\end{enumerate}   
The latter effect is only understood when taking into account gradient and curvature drifts: Cosmic ray proton spectra are softer during an A$>$0-cycle so that below 500~MeV the A$>$0-solar minima spectra are always higher than the corresponding A$<$0-spectra 
\replaced{[Beatty et al., 1985; Potgieter and Moraal, 1985]}{\citep{Kota-Jokipii-1983,Beatty-etal-1985,Potgieter-Moraal-1985}}. This means that the adiabatic energy losses that cosmic rays experience are somewhat different in both cycles \citep{Strauss-etal-2011}, and also cause the proton spectra for two consecutive solar minima to cross at a few GeV \citep{Reinecke-Potgieter-1994}. 
Thus, it is questionable to apply the commonly used energy spectra, derived from neutron monitor measurements utilizing the force field solution, to phenomena that are predominantly caused by lower energy ions. 
Recently, \citet{Corti-etal-2016} performed a similar approach as laid out%
\deleted{by Gieseler et al. [2015] and}%
in this work, utilizing 
a force field modification with two $\phi$ parameters to describe GCR spectra measured over a large energy range by BESS, PAMELA, and AMS-02 but only for single selected time periods from 1993 to 2011. 
\citet{Cholis-etal-2016} presented a rigidity dependent modulation parameter which is not derived from direct GCR measurements but instead of measurements of the heliospheric current sheet tilt angle and the magnetic field amplitude. 

In what follows we will briefly recall the derivation of the force field solution following the work by \citet{Moraal-2013}. In the following section it is shown that helium and carbon measurements at 1.27~GV aboard IMP-8 and ACE, respectively, are a good proxy for the temporal variation of protons at the same rigidity. Normalizing these count rates to the proton intensities at 1.28~GV measured by PAMELA, we can derive a solar modulation potential at these rigidities. 
Following the arguments from \citet{Herbst-etal-2010,Herbst-etal-2017}, this solar modulation potential, however, does not necessarily have to be in agreement with the one derived by \citet{Usoskin-etal-2005}, \citet{Usoskin-etal-2011} but also not with the one given by \citet{Gil-etal-2015} or \citet{Usoskin-etal-2017} because it is derived at a different rigidity range. 
In order to ascribe the full spectrum, we perform a detailed analysis of the rigidity dependence of the solar modulation potential using the high-precision PAMELA measurements, and show that a weighted combination of two modulation potentials is capable to describe the rigidity spectra for a full Hale cycle.

\section{Cosmic ray transport in the heliosphere}
\label{section:Cosmic ray transport in the heliosphere}
The transport of cosmic rays inside the heliosphere was first described by 
\citet{Parker-1965}:
\begin{align}
\frac{\partial f}{\partial t} = 
 - \big(\underbrace{\textbf{V}}_\mathrm{i} + \underbrace{\langle {\bf v}_D\rangle}_{\mathrm{ii}} \big)\cdot \nabla f 
+  \underbrace{\nabla \cdot \left({\bf \stackrel{\leftrightarrow}{\kappa}}
\cdot \nabla f\right)}_{\mathrm{iii}} 
 + \underbrace{\frac{1}{3}(\nabla \cdot {\bf V})
\frac{\partial f}{\partial \ln P}}_{\mathrm{iv}} +  \underbrace{Q}_{\mathrm{v}},
\label{eq:parker}
\end{align}
where $f({\bf r},P,t)$ denotes the differential cosmic ray phase space distribution function, $\textbf{r}$ the spatial coordinates, $P$ the particle rigidity, $t$ the time, and
\begin{itemize}
\item[$\mathrm{i})$] the outward convection by the solar wind speed $\textbf{V}$,
\item[$\mathrm{ii})$] the gradient and curvature drifts in the global heliospheric magnetic field \citep{Jokipii-etal-1977},
\item[$\mathrm{iii})$] the diffusion through the irregular heliospheric magnetic field,
\item[$\mathrm{iv})$] the adiabatic energy change due to the divergence of the expanding solar wind, and
\item[$\mathrm{v})$] the local sources like particles accelerated at the Sun.
\end{itemize}
Although the modulation of GCRs in the heliosphere strongly depends on all the processes mentioned above, a much simpler analytical approximation can be derived from Eq.~\ref{eq:parker}. Following \citet[][see also \citet{Gleeson-Axford-1968, Caballero-Lopez-Moraal-2004}]{Moraal-2013}, Eq.~\ref{eq:parker} can be reduced to a simple convection-diffusion equation if there is 
(a) no source of cosmic rays ($Q = 0$), 
(b) a steady state (${\partial f}/{\partial t} = 0$), 
(c) an adiabatic energy loss rate $\langle {dP}/{dt}\rangle = (P/3)\textbf{V}\cdot \nabla f/f=0$, 
and (d) no drifts. 
Assuming spherical symmetry, i.~e. only the radial direction is taken into account, this leads to
\begin{align}
\frac{v P}{3}\frac{\partial f}{\partial P} + \kappa \frac{\partial f}{\partial r} = 0, \label{eq:conv-diff}
\end{align}
with $v = V_r$ denoting the solar wind speed. If the diffusion coefficient $\kappa(r,P)$ is separable $\kappa=\kappa_1(r)\cdot \kappa_2(P)$ with $r$ the heliocentric distance and $P$ the particle rigidity, and furthermore $\kappa_2(P) \propto P$, the following expression for the so-called force field parameter (or solar modulation potential) $\phi$ can be obtained:
\begin{align}
\phi(r) = \int\limits_r^{r_b} \frac{v(r')}{3\kappa_1} dr'.
\end{align}
Here $r_b$ represents the outer boundary like the solar wind termination shock or the heliopause (cf. \citet{Caballero-Lopez-Moraal-2004}). Typical modulation values, still depending on the LIS model used, vary between 300 and 1500~MV with increasing solar activity. 
\citet{Gleeson-Urch-1973} as well as \citet{Caballero-Lopez-Moraal-2004} and \citet{Moraal-2013} investigated the validity of the force field approximation by comparing its results with a full numerical solution of the steady state, spherically symmetric (one-dimensional) transport equation
and direct measurements (see e.~g. Fig.~5 in \citet{Moraal-2013}). Although they found that the approximation starts to deviate from the full numerical solution at energies below $\sim$150-550~MeV and when going to the outer heliosphere, it is still a useful way to describe differential intensity spectra $J_{\mathrm{1AU}}$ at 1~AU during intermediate and low solar activity by using the following equation:
\begin{align}
J_{\mathrm{1AU}}(E,\phi) = J_{LIS}(E+\Phi)\frac{(E)(E+2 E_r)}{(E+\Phi)(E+\Phi+2 E_r)}
\label{eq:force-field-equation}
\end{align}
The force field function $\Phi$ is given by $\Phi = (Z e/A)\phi$, where $Z$ and $A$ are the charge and mass number of the cosmic ray nuclei, respectively. $E$ represents the kinetic energy of the particles, $E_r$ their rest energy ($E_r=0.938$~GeV for protons) and $J_{LIS}$ gives the differential energy spectra of the LIS representing the boundary condition of the force field approximation. However, the full LIS by now has not been measured, thus multiple LIS-models exist in the literature.

In what follows, we perform a $\chi^2$ minimization process similar to \cite{Wiedenbeck-et-al-2005} to derive the solar modulation potential $\phi$ for an actual measurement of the intensity spectrum $J_{\mathrm{1AU}}$ at 1~AU. First, we generate model intensity spectra for the investigated energy (respectively rigidity) range using the force field solution with varying $\phi$. In this process we use either the LIS from \cite{Burger-etal-2000} as described by \cite{Usoskin-etal-2005} (used for a more in-detail analysis later on) or the newer model by \citet{Vos-Potgieter-2015}, which is used by \citet{Usoskin-etal-2017}. Then, for each spectrum we calculate the sum-of-squares deviation to the measured spectrum, and choose that $\phi$ with the smallest deviation.

\section{Observation and data analysis}

\label{section:observation}
As already mentioned, the energy (rigidity) dependent modulation of galactic cosmic rays (GCRs) with solar activity is shown in the upper panel of Fig.~\ref{fig:jgr_icrc2015_sunspot}, where the intensity variations of the Kiel neutron monitor (black curve) and of 1.28~GV proton proxies (blue curves) as a proxy for high and low energy GCRs are plotted over time, respectively. A simple comparison with the sunspot number in the panel below gives the anti-correlation between solar activity and GCR intensity. 
As described in Sect.~\ref{section:Cosmic ray transport in the heliosphere}, the time profile of high energy GCRs in the inner heliosphere can be reasonably approximated by the force field approximation. In this process, the energy spectrum of a GCR species at 1~AU is derived from its unmodulated local interstellar spectrum (LIS) only by the modulation potential $\phi$.

\subsection{Proton proxies}
\label{section:proton-proxies}
Neutron monitors at Earth have been proven to be very reliable proxies for long-time GCR measurements. However, they are limited with respect to the observable energies due to the shielding of the Earth's magnetic field and atmosphere. To measure energies below the GeV range one has to take advantage of balloon-borne or spacecraft experiments. In this work we will use energetic particle observations from the spacecraft ACE, IMP-8, and PAMELA to cover the time period from the 1970s to the last, commonly called unusual solar minimum in 2009 and beyond.

\begin{figure}
\includegraphics[width=\columnwidth]{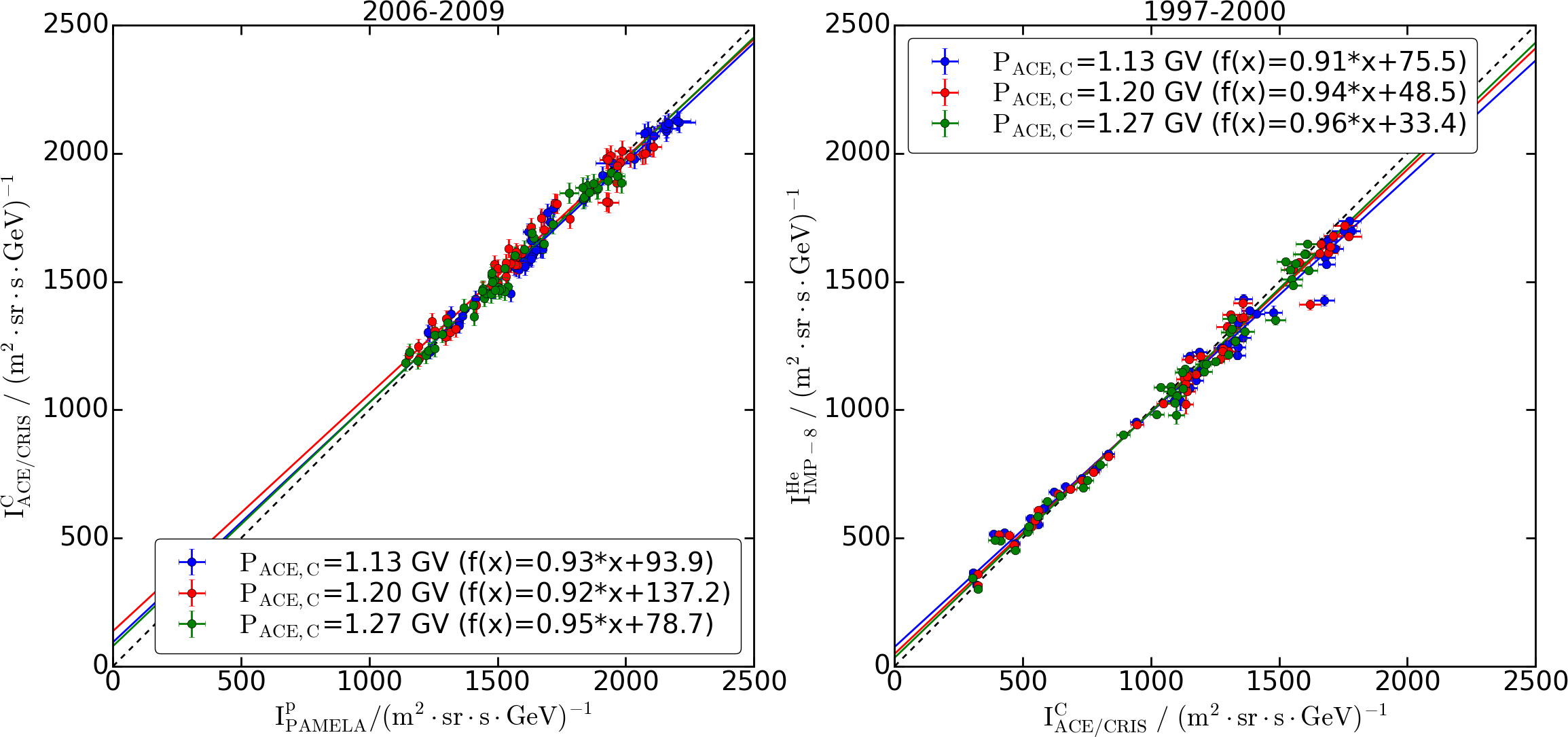}
\caption{Left: Normalized carbon intensities from ACE/CRIS with respect to PAMELA proton intensities at three corresponding rigidities. Right: Normalized helium intensities from IMP-8 (1.03-1.45~GV) with respect to normalized ACE/CRIS carbon intensities at three different rigidities. (All data with statistical errors.) For each data set the best fit linear regression is given.}
\label{fig:proton_proxy_norm}
\end{figure}
In a first step we are interested in the temporal behavior of different ions with the same rigidity. 
The rigidity $P$ is calculated from the particle momentum $p$ by $P = \frac{pc}{|Z|e}$. 
It has been shown previously that ions with the same ratio $\frac{Z e}{A}$ like helium, carbon or oxygen undergo the same temporal variation \citep[e.g.][]{Webber-etal-2005,Gieseler-etal-2008,Heber-etal-2008,McDonald-etal-2010}. 
We take advantage of this and use {IMP-8} helium and ACE/CRIS carbon measurements as proxies for protons at the same rigidity. 
Figure~\ref{fig:proton_proxy_norm} shows the correlation between monthly proton intensities measured by PAMELA \citep{Adriani-etal-2011a} and normalized ACE/CRIS carbon intensities (ACE Science Center) at three corresponding rigidities (the highest available for ACE/CRIS carbon) in the time 2006-2009 (left), and the same three carbon intensities correlated with normalized {IMP-8} helium intensities (F.~B. McDonald, private communication) at 1.03-1.45~GV for 1997-2000 (right), respectively. In both cases the three different ACE/CRIS carbon intensities show good linear correlations with the PAMELA protons and IMP-8 helium, yielding slopes from $0.93\pm0.02$ to $0.95\pm0.02$ and $0.91\pm0.01$ to $0.96\pm0.02$ for the best fit linear regressions, respectively. 

\begin{figure}
\includegraphics[width=1\columnwidth]{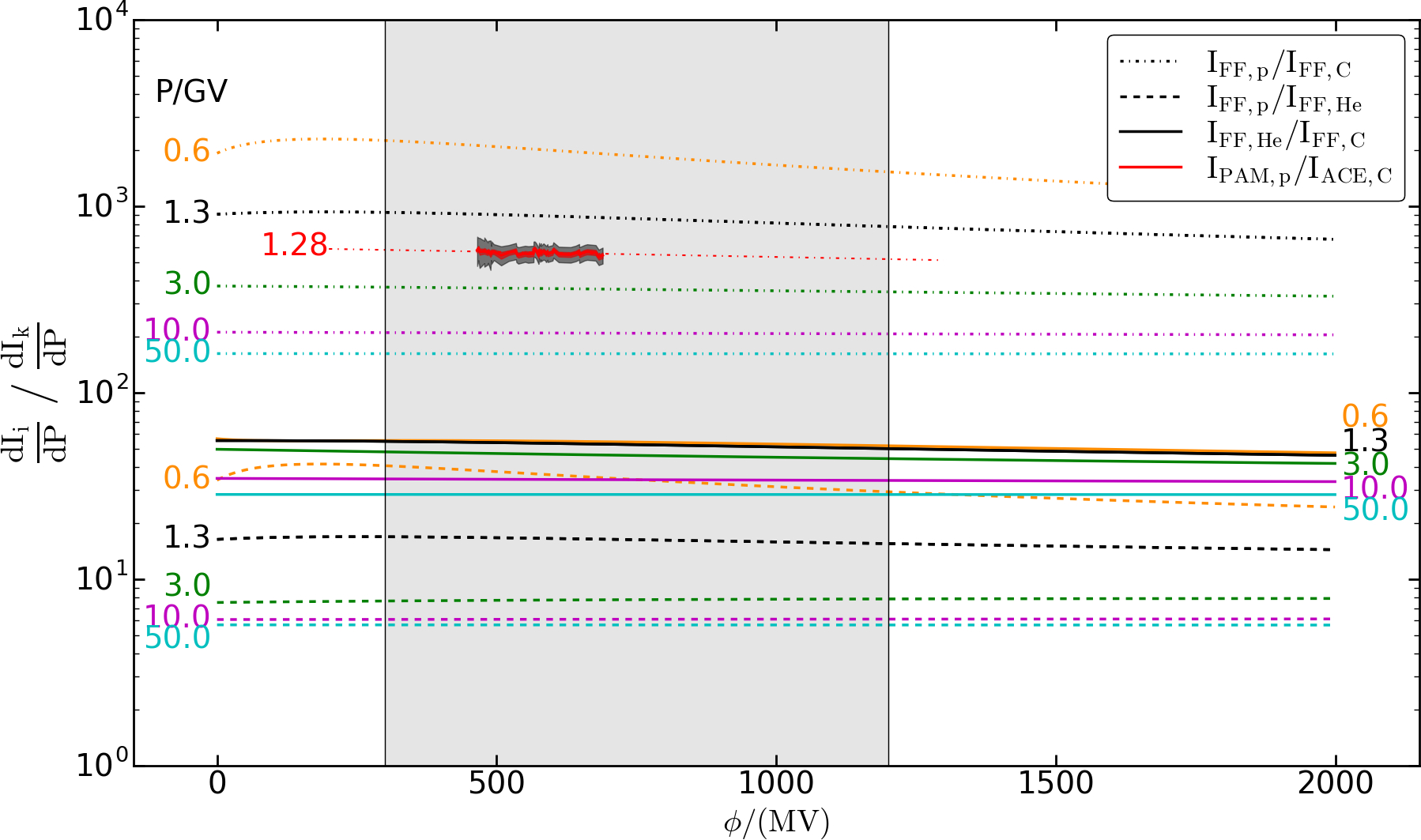}
\caption{Dependency of the ratio of force field model differential rigidity spectra of protons to carbon (dashed-dotted lines), helium to carbon (solid lines), and proton to helium (dashed lines), respectively, with respect to the solar modulation potential $\phi$ and for different rigidities (denoted next to lines). All LIS based on \citet{Bishoff-etal-2016}. The red line shows the ratio of the proton intensities measured by PAMELA at 1.28~GV and the carbon intensities measured by ACE/CRIS at 1.27~GV with statistical errors (gray shaded area with additional systematic errors for PAMELA; red dot-dashed line shows a fit line through the measurements). Marked by shading from 300 to 1200~MV is the range of typical solar modulation.
}
\label{fig:int_ratio_wrt_phi_rig_biss}
\end{figure}

Although the different particles should undergo the same temporal variations by the solar modulation, differences in the spectral slopes of their individual LIS can lead to variations in their intensity ratios. The significance of this effect has been investigated by calculating the differential rigidity spectra ratios at Earth of protons to carbon, helium to carbon, and protons to helium, respectively. Here the force field approach with different LIS for all three particles has been used. We choose to utilize the model from \citet{Bishoff-etal-2016} because it provides independent LIS for each investigated species: proton, helium and carbon.
In Fig.~\ref{fig:int_ratio_wrt_phi_rig_biss}, these ratios are plotted with respect to the solar modulation potential for different rigidities. 
At low particle rigidities all ratios show a dependency with the solar modulation potential, which gets significant for very low rigidities. At the observed 1.3~GV the intensity ratios can vary by a maximal factor of 19.2\% for protons to carbon, 8.9\% for protons to helium, and 9.5\% for helium to carbon, respectively, when comparing intensities at low ($\phi$=300~MV) and high ($\phi$=1200~MV) solar activity. However, the effect vanishes when intensities are compared at the same level of solar activity.
For the periods with solar modulation potentials between 530~MV and 690~MV (i.~e. years 2006-2009) the ratio of the measured 1.28~GV proton intensity from PAMELA and 1.27~GV carbon intensity from ACE/CRIS follows the same behavior, albeit with a constant offset.
In what follows we will use the 1.27~GV ACE/CRIS carbon intensity, normalized to the 1.28~GV PAMELA proton intensity, and the 1.03-1.45~GV IMP-8 helium intensity, normalized to the normalized 1.27~GV ACE/CRIS carbon intensity, as proxies for 1.28~GV protons.

\subsection{Intensity time profiles}
\label{section:intensity-time-profiles}
\begin{figure*}
\includegraphics[width=\linewidth]{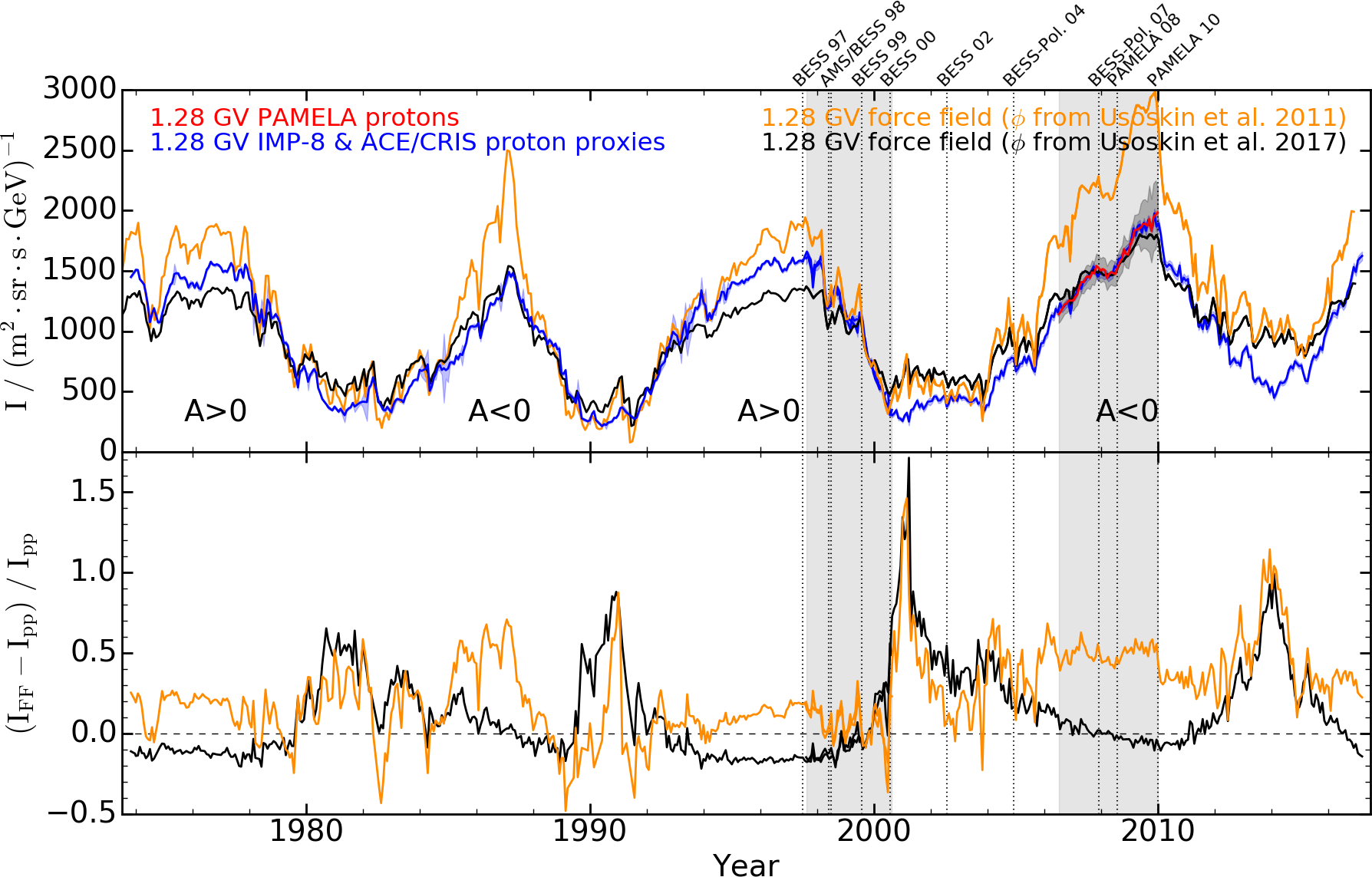}
\caption[]{
Top: Measured and calculated intensity time profiles of 1.28~GV protons. The red and blue curves reflect the measurements of protons by PAMELA (with statistical uncertainties in red and systematics given by gray shaded area, respectively) as well as proton proxies by {IMP-8} helium (1973-2000) and ACE/CRIS carbon (1997-2017, both with statistical errors), respectively.
The calculated intensity of the force field solution at 1.28~GV using the modulation potential from \citet{Usoskin-etal-2011} based on the LIS from \citep{Burger-etal-2000}
is displayed by the orange curve while the black curve shows the same intensity for the updated modulation potential as given by \citet{Usoskin-etal-2017} using the LIS from \citep{Vos-Potgieter-2015}.
Bottom: The deviation of the calculated force field intensities for $\phi$ from \citet{Usoskin-etal-2011} (orange) and \citet{Usoskin-etal-2017} (black) compared to the proton proxies, respectively. 
The two shaded time periods mark the normalization interval of {IMP-8} helium to ACE/CRIS carbon and of ACE/CRIS carbon to PAMELA protons, respectively (cf. Fig.~\ref{fig:proton_proxy_norm}).
Vertical dotted lines with annotations above plot indicate the measurement periods used in Fig.~\ref{fig:verification}.
}
\label{fig:jgr_icrc2015_pamuso}
\end{figure*}

Figure~\ref{fig:jgr_icrc2015_pamuso} (top) displays the intensity time profiles of 1.28~GV protons measured by PAMELA (red curve), and the corresponding proton proxies derived from {IMP-8} helium (1973-2000, blue curve) and ACE/CRIS carbon measurements (1997-2017, blue curve). 
The orange curve shows the intensity time profile at 1.28~GV using the solar modulation potential from \citet{Usoskin-etal-2011} based on the LIS from \citep{Burger-etal-2000}  while the black curve shows the same intensity for the updated modulation potential as given by \citet{Usoskin-etal-2017} using the LIS from \citep{Vos-Potgieter-2015}.
The bottom panel of Fig.~\ref{fig:jgr_icrc2015_pamuso} shows the deviation of the calculated force field intensities for $\phi$ from \citet{Usoskin-etal-2011} (orange) and \citet{Usoskin-etal-2017} (black) compared to the proton proxies, respectively.
Utilizing the solar modulation values from \citet{Usoskin-etal-2011} the intensities are overestimated by up to 70\% and 50\% during the 1980's and 2000's solar minima, respectively. The differences are significantly smaller during the 1970's and 1990's. This behavior is somehow expected due to the hardening of the spectra during an A$<$0-solar magnetic epoch.
\citet{Usoskin-etal-2017} presented an updated version of their modulation potential reconstruction (for the energy range 1-30~GeV, i.~e. 1.7-30.9~GV), which includes three main changes: (1) the usage of the new yield function for neutron monitors at sea level by \citet{Mishev-etal-2013}; (2) the usage of a recent LIS by \citet{Vos-Potgieter-2015}, which incorporates Voyager measurements in the outer heliosheath; and (3) the calibration of the neutron monitor response to the direct PAMELA proton measurements from 2006-2009. Because of the changes in the used LIS, the solar modulation potentials from \citet{Usoskin-etal-2005,Usoskin-etal-2011} and \citet{Usoskin-etal-2017} can not be compared directly (cf. Fig.~10 in \citet{Usoskin-etal-2017}). However, the yielded intensities are comparable and shown in Fig.~\ref{fig:jgr_icrc2015_pamuso} (top). 
The intensity calculated using the force field solution with the new modulation potential from \citet{Usoskin-etal-2017} (black line) agrees quite well with the PAMELA proton intensities from 2006-2009 because of the calibration to these measurements, and also with the proton proxies during the previous A$<$0-cycle (cf.~Fig.~\ref{fig:jgr_icrc2015_pamuso} (bottom)). \citet{Usoskin-etal-2017} noted that their model may slightly underestimate the modulation during periods of high solar activity (i.~e. low solar modulation potential). This can be seen in the time periods around 1980-1983, 1989-1992, 2001-2004, and 2012-2016, where their model overestimates the intensities compared to the measured proton proxies by up to 50\%, 85\%, 130\% and 100\%, respectively. However, more important are the discrepancies between the \citet{Usoskin-etal-2017} model and the proton proxies in the $A>0$-cycles. During the 1970's and 1990's solar minima the model underestimates the intensities permanently by up to 20\%. Note that this cannot be explained by the effect of different LIS for protons, helium and carbon (as investigated in Fig.~\ref{fig:int_ratio_wrt_phi_rig_biss}) because at all solar minima there were comparable intensity levels and thus $\phi$-values. The only exception is in 2009 where the highest space-age intensities were detected, but this time period does not show any different behavior.

\subsection{Solar modulation potential for different rigidity ranges}
\label{section:solar-modulation-potential-for-different-rigidity-ranges}

\citet{Gil-etal-2015} and \citet{Usoskin-etal-2017} showed that the solar modulation potential $\phi$ from \citet{Usoskin-etal-2011}, which is calculated using neutron monitor observations, is not in agreement with $\phi$-values obtained by the analysis based on PAMELA proton data. 
Details of the fitting procedure of the PAMELA data are only given in \citet{Usoskin-etal-2017}, where $\phi$ is fitted to proton energies from 1 to 30~GeV (i.~e. 1.7 to 30.9~GV), which the authors claimed is the most effective part of the energy spectrum for GCR detection by neutron monitors. 
In the following we investigate in more detail which part of the PAMELA proton spectra should be used to calculate the solar modulation potential in order to compare it with one derived by neutron monitor measurements.
\begin{figure}
\includegraphics[width=\columnwidth]{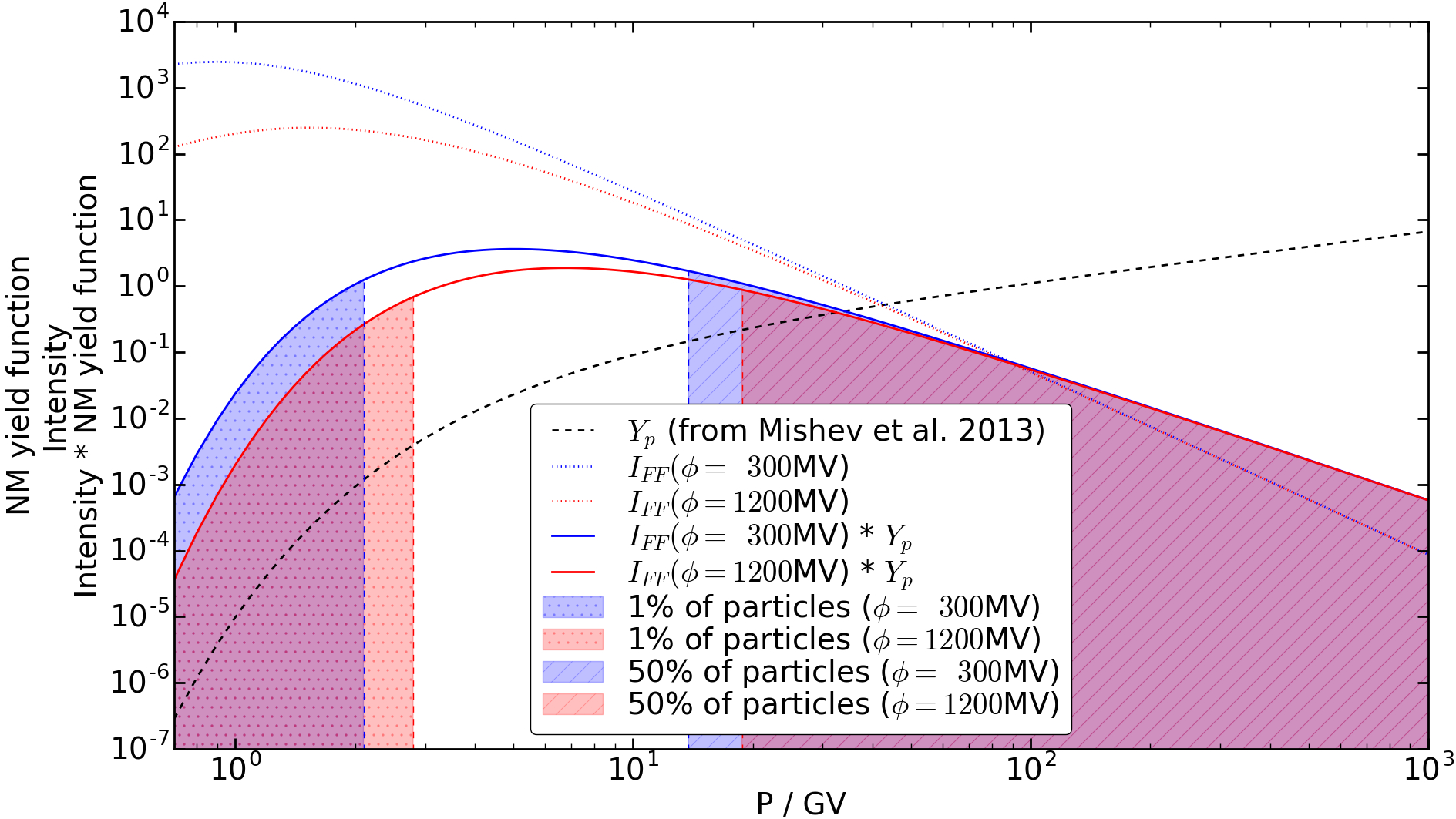}
\caption{Rigidity-dependent force field proton intensity spectra $I_{FF}$ for two typical solar modulation potentials (300 and 1200~MV) folded with the proton yield function $Y_p$ from \citet{Mishev-etal-2013} (arbitrary units on y-axis). Marked by blue and red shading are the areas containing 1\% and 50\% of the particles, respectively.}
\label{fig:nm_yield_print_0_01}
\end{figure}
\begin{figure}
\includegraphics[width=\columnwidth]{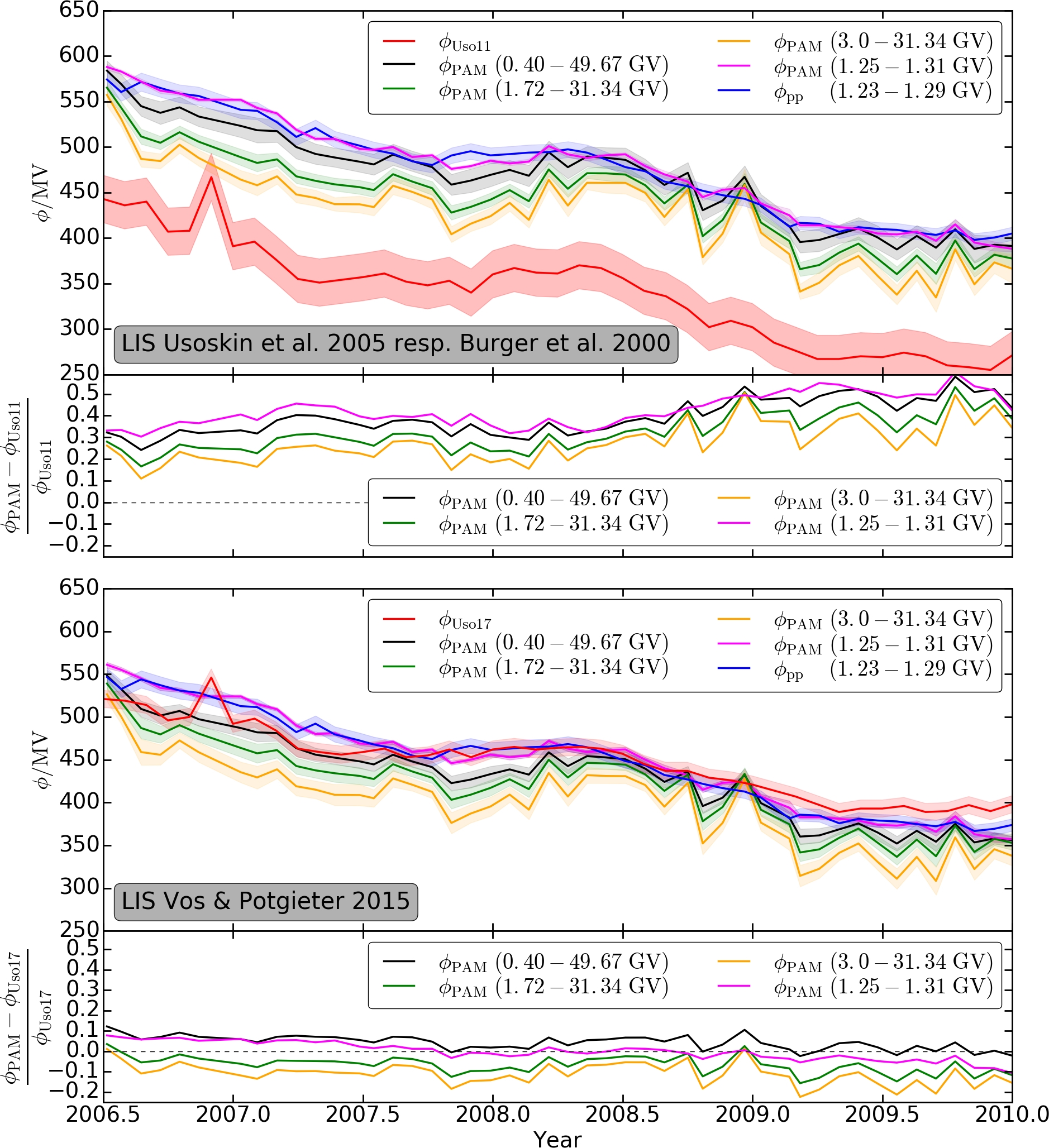}
\caption[]{Top two panels: Solar modulation potential (with uncertainty range) from \cite{Usoskin-etal-2011} based on neutron monitor measurements (red curve), derived from 1.23-1.29~GV proton proxies (ACE/CRIS carbon, blue curve) and from PAMELA proton measurements in different rigidity ranges. All modulation potentials were calculated using the LIS from \citet{Burger-etal-2000}. The deviations of the solar modulation potentials derived by PAMELA to those from \cite{Usoskin-etal-2011} are plotted in the panel below.
Bottom two panels: Same as above, but now the solar modulation potentials are calculated using the LIS from \citet{Vos-Potgieter-2015} or are taken from \cite{Usoskin-etal-2017}, respectively.
}
\label{fig:phi_comparison_2006_2010_bu_vos}
\end{figure}
As discussed in Sect.~\ref{section:introduction}, the most recent yield function of a sea level neutron monitor is given by \citet{Mishev-etal-2013}. Their Fig. 3 indicates the strong decrease of the yield function with decreasing energies below 10~GeV.  
To investigate the implications of this effect, we used the force field approach to generate rigidity-dependent GCR proton intensity spectra for two typical solar modulation potentials, 300 and 1200~MV, reflecting low and high solar activity, respectively. These spectra are then folded with the proton yield function from \citet{Mishev-etal-2013} \citep[in the analytical form given by][]{Caballero-Lopez-2016}, resulting in Fig.~\ref{fig:nm_yield_print_0_01}. 
Marked by shading are the areas containing 1\% and 50\% of the particles; i.~e. the rigidity range up to 3~GV contains only 1\% of all particles for the scenario with $\phi=1200$~MV (and even less for a smaller $\phi$), while up to 15-20~GV 50\% of the particles are detected, yielding an approximation for the mean neutron monitor rigidity.
This indicates that neutron monitors are almost not sensitive to the rigidity range below 3~GV, where the PAMELA proton data shows the strongest modulation effects.

\begin{figure}
\includegraphics[width=\columnwidth]{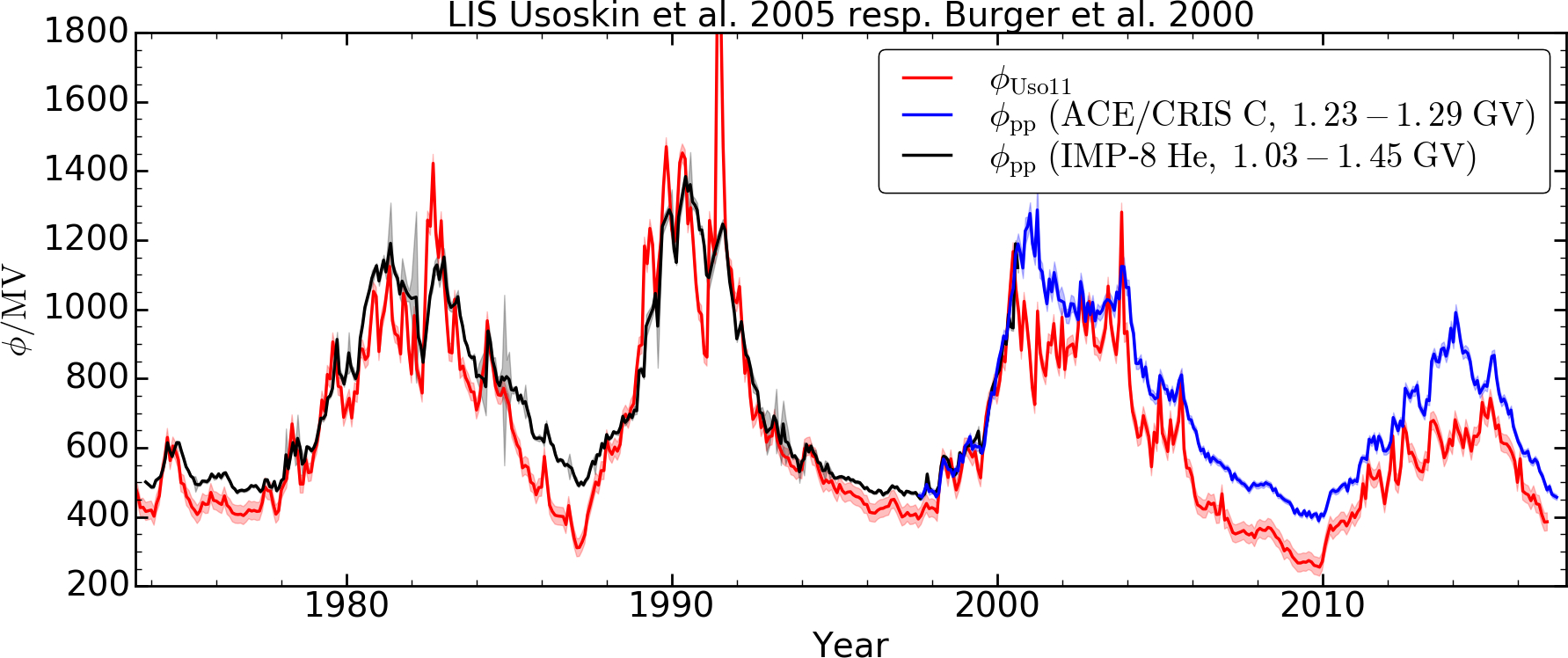}
\caption[]{
Similar to Fig.~\ref{fig:phi_comparison_2006_2010_bu_vos} (top) but now for the time period 1973-2017:
Solar modulation potential from \cite{Usoskin-etal-2011} based on neutron monitor measurements (red curve) and derived from 1.28~GV proton proxies (IMP-8 helium and ACE/CRIS carbon, black and blue curve, respectively).
All modulation potentials were calculated using the LIS from \citet{Burger-etal-2000}. 
}
\label{fig:phi_comparison_1973_2017}
\end{figure}

The top panel of Fig.~\ref{fig:phi_comparison_2006_2010_bu_vos} shows the solar modulation potential during the time period 2006-2010 derived as described in Sect.~\ref{section:Cosmic ray transport in the heliosphere} using the proton proxies and different rigidity ranges of the PAMELA proton measurements together with the neutron monitor based results of \citet{Usoskin-etal-2011} ($\phi_{Uso11}$), and the deviations between these findings. All potentials were calculated using the LIS of \citet{Burger-etal-2000}. 
Figure~\ref{fig:phi_comparison_2006_2010_bu_vos} (bottom) shows the same potentials from proton proxies and PAMELA as well as results from \citet{Usoskin-etal-2017} ($\phi_{Uso17}$) but now the LIS of \citet{Vos-Potgieter-2015} was used.
$\Delta\phi_{Uso11}$ is given as a %
$1\sigma$ uncertainty of 26~MV \citep{Usoskin-etal-2011}, and $\Delta\phi_{Uso17} < 10$~MV \citep[][in Fig.~\ref{fig:phi_comparison_2006_2010_bu_vos} (bottom) 10~MV is shown]{Usoskin-etal-2017}.
The uncertainties $\Delta\phi_{pp}$ of the calculated solar modulation potential derived from the proton proxy measurements are determined by estimating the influences of the uncertainties of rigidity, $\Delta P$, and of intensity, $\Delta I$, as followed:
\begin{align}
\Delta\phi_{P}  & = \max \left|\phi(P,I)-\phi(P\pm\Delta P,I)\right| 
\label{eq:delta-phi-p}\\
\Delta\phi_{I}  & = \max \left|\phi(P,I)-\phi(P,I\pm\Delta I)\right|
\label{eq:delta-phi-i}\\
\Delta\phi_{pp} & = \sqrt[]{{\Delta\phi_{P}}^2 + {\Delta\phi_{I}}^2}
\label{eq:delta-phi}
\end{align}
Here $\Delta I$ is the statistical uncertainty of the measured intensities. $\Delta P$, the uncertainty of the mean rigidity of the single proton proxy measurement channel, is estimated by using a force field intensity $I_{FF}$ to calculate the spectral-dependent mean rigidity at very low ($\phi=200$~MV) and at very high solar activity ($\phi=1500$~MV) following:
\begin{align}
\langle P \rangle & = \frac{\sum I_{FF}(P_i,\phi)\cdot P_i}{\sum I_{FF}(P_i,\phi)}\label{eq:delta-p}\\
\Delta P & = \frac{1}{2}\cdot \left|\langle P_{low}\rangle -\langle P_{high}\rangle\right|
\end{align}
The different $\phi_{\textrm{PAM}}$ are derived from different parts of the PAMELA proton spectrum by a non-linear least squares fit procedure weighting each data point by its uncertainties in rigidity and intensity. To estimate their errors $\Delta\phi_{\textrm{PAM}}$ the $1\sigma$ uncertainties are used.  
An exception is $\phi_{\textrm{PAM}}$ for the single channel PAMELA proton measurement '1.25-1.31~GV'. Because it contains only one rigidity data point, the same approach as for the $\Delta\phi_{pp}$ is used here.

The significant difference between $\phi_{Uso11}$ and the potentials from PAMELA, $\phi_{PAM_i}$, is visible in Fig.~\ref{fig:phi_comparison_2006_2010_bu_vos} (top), especially if it is compared to the potentials derived from single measurement channels at lower rigidities (magenta and blue lines) or from the whole spectrum from 0.4 to 50~GV (black line).  
This offset gets smaller when only higher rigidity parts of the PAMELA proton spectrum are used for the potential calculation, starting from the rigidity range 1.7-31.3~GV (green line) used by \citet{Usoskin-etal-2017} up to neglecting all lower rigidity particles and also the high rigidities where almost no solar modulation takes place (3-31.3~GV, orange line). 
Still the offset is between 10\% and 50\%. Furthermore, the deviations of $\phi_{PAM_i}$ to $\phi_{Uso11}$ show a small temporal trend. Altogether, from Fig.~\ref{fig:phi_comparison_2006_2010_bu_vos} (top) it seems obvious that the smallest deviation can be achieved if the lower rigidities are left out from the $\phi$ calculation, as indicated by Fig.~\ref{fig:nm_yield_print_0_01}, and only the spectrum from 3~GV upwards is used. Defining this threshold at higher rigidities would result in losing sensitivity to the solar modulation, which significantly decreases with increasing rigidity in this range. This will be shown in more detail in Sect.~\ref{section:rigidity-dependence-of-the-solar-modulation-potential}.
The overall picture seems to be more coherent in Fig.~\ref{fig:phi_comparison_2006_2010_bu_vos} (bottom), where $\phi_{Uso17}$ was calculated using PAMELA measurements to calibrate the neutron monitor responses in the investigated time interval. This results in much smaller deviations between $\phi_{PAM_i}$ and $\phi_{Uso17}$. But also in the fact that the neutron monitor based $\phi_{Uso17}$ now better reflects the lower rigidity parts of the PAMELA proton spectrum than the rigidity ranges which should correspond to neutron monitor measurements. 

Because we want to use a solar modulation potential reflecting the neutron monitor measurements in the following analysis, we continue using $\phi_{Uso11}$ from \citet{Usoskin-etal-2011} and calculate $\phi_{pp}$ reflecting the proton proxy measurements using the LIS from \citet{Burger-etal-2000} for the time period from 1973 up to 2017. These results are presented in Fig.~\ref{fig:phi_comparison_1973_2017} and 
are available as data set S1 in the supporting information (also online at \url{http://www.ieap.uni-kiel.de/et/ag-heber/cosmicrays}).

\subsection{Rigidity dependence of the solar modulation potential}
\label{section:rigidity-dependence-of-the-solar-modulation-potential}
\begin{figure}
\includegraphics[width=\columnwidth]{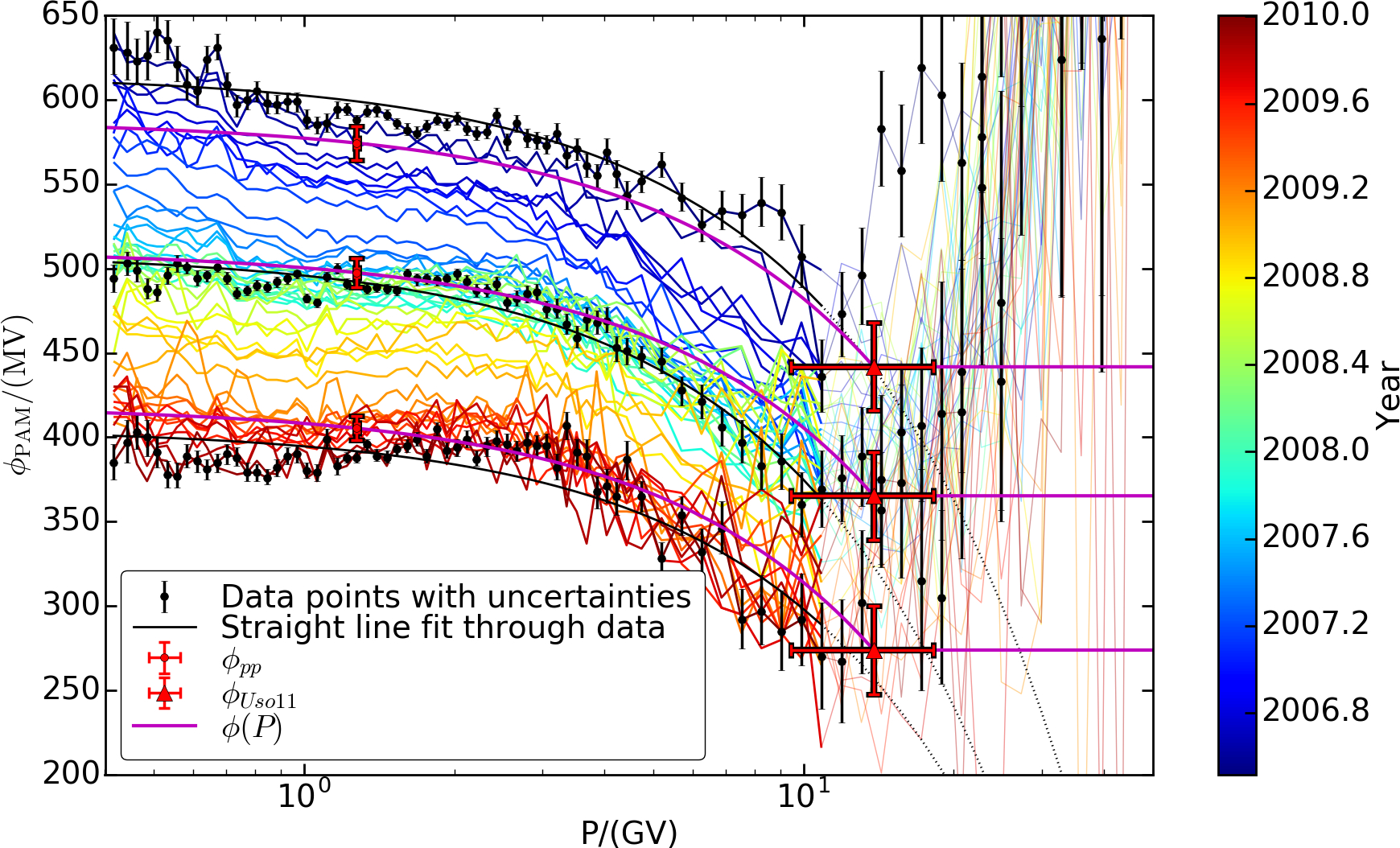}
\caption[]{
Solar modulation potential derived from PAMELA proton measurements for each rigidity bin and monthly measurement interval (colored lines). The values above 10~GV (plotted with lighter colors) are not reliable, see text for details.
In addition, for the first, mid and last investigation period all data points with uncertainties are plotted (black points). Through each ensemble of this data points a straight line fit is calculated (in linear space), omitting the range above 10~GV (black solid line; the dotted line prolongs this fit to higher rigidities). 
For these three periods also the measured $\phi_{pp}$ and $\phi_{Uso11}$ with their uncertainties are included (red data points). Assuming a linear relationship, each pair of $\phi_{pp}$ and $\phi_{Uso11}$ is connected by a straight line (in linear space), yielding the rigidity-dependent $\phi(P)$ as given by Eq.~\ref{eq:rig-dep-phi} (magenta line). 
}
\label{fig:pam_phis_wrt_rigidity_bu_detail2}
\end{figure}
In order to derive the rigidity dependence of the solar modulation potential we utilize the monthly averaged proton measurements from PAMELA \citep{Adriani-etal-2011a} and apply the following procedure:
\begin{enumerate}
\item The LIS from \citet{Burger-etal-2000} is used as the input spectrum for Eq.~\ref{eq:force-field-equation}.
\item For each small PAMELA rigidity bin the solar modulation potential $\phi_{PAM_i}$ that fits best the measured intensity is determined by a minimizing process.
\item These solar modulation potentials are plotted with respect to rigidity (colored lines in Fig.~\ref{fig:pam_phis_wrt_rigidity_bu_detail2}).
\end{enumerate}
Since the modulation is small compared to the measurement uncertainties at rigidities above 10~GV, the minimization process is not reliable here and the yielding $\phi$-values are not representative. Note that this picture also depends on the used LIS model. 
Fig.~\ref{fig:pam_phis_wrt_rigidity_bu_detail2} clearly indicates a non-constant dependency of the solar modulation potential with respect to rigidity. This shows that it is not reasonable to describe GCR intensities in the inner heliosphere by only one rigidity-independent parameter $\phi$. 
To get an analytically description of the rigidity dependence of $\phi_{PAM}$ straight line fits have been calculated for each monthly averaged proton measurement, omitting the unreliable rigidity range above 10~GV (solid black lines in Fig.~\ref{fig:pam_phis_wrt_rigidity_bu_detail2}). These fits are used to calculate the corresponding rigidities for the neutron monitor-based solar modulation potentials $\phi_{Uso11}$: For each monthly potential the rigidity at which the corresponding fit line has the same value is calculated, afterwards the mean and standard deviation of these rigidities is calculated. 
This yields a mean rigidity for $\phi_{Uso11}$ of $P_{Uso11}=13.83\pm4.39$~GV, which is in the rigidity range expected from Fig.~\ref{fig:nm_yield_print_0_01}. With these two solar modulation potentials for two different rigidities in the spectrum we can now establish a full rigidity-dependent modulation potential.

\section{The two parameter force field approach}
\label{section:two-parameter-force-field-approach}
\begin{figure}
\includegraphics[width=\columnwidth]{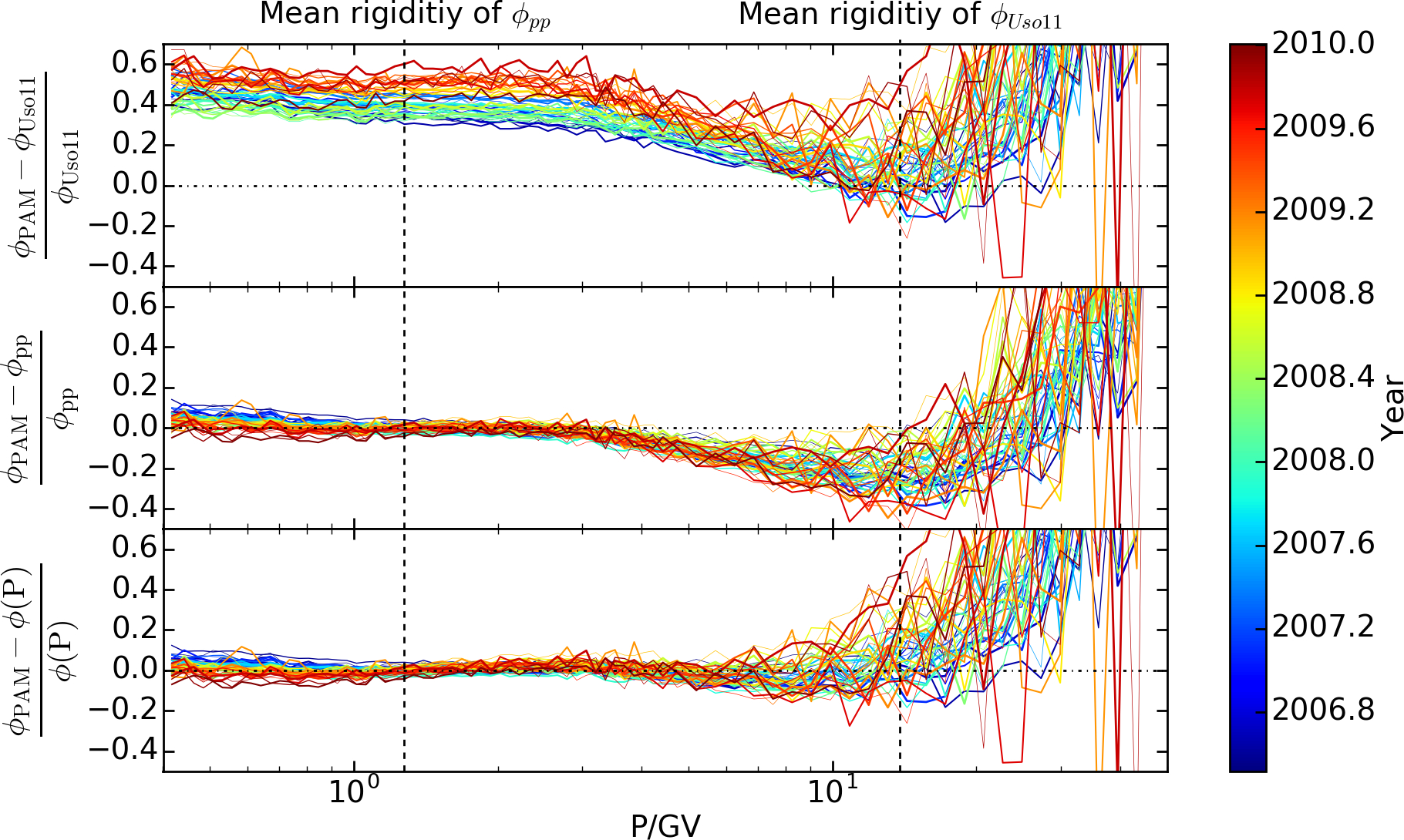}
\caption[]{
Deviations of the solar modulation potential $\phi_{PAM}$, derived from PAMELA proton measurements for each rigidity bin and monthly measurement interval (as in Fig.~\ref{fig:pam_phis_wrt_rigidity_bu_detail2}), to the rigidity-independent $\phi_{Uso11}$ (top) and $\phi_{pp}$ (middle) as well as the rigidity-dependent $\phi(P)$ (bottom), respectively.
}
\label{fig:pam_phis_wrt_rigidity_ratios}
\end{figure}
In Sect.~\ref{section:rigidity-dependence-of-the-solar-modulation-potential} 
we showed that it is not sufficient to describe GCR intensities at Earth by only one rigidity-independent parameter. As a workaround, we now present a modified force field approach which utilizes two solar modulation potentials at two different rigidities. We use these two parameters:
\begin{enumerate}
\item $\phi_{pp}$, derived from the 1.28~GV proton proxies IMP-8 helium and ACE/CRIS carbon, which are normalized to PAMELA proton measurements at the same rigidity.
\item $\phi_{Uso11}$, calculated by \citet{Usoskin-etal-2011} and representing the neutron monitor measurements at higher rigidities. 
\end{enumerate}
We base our analysis on the modulation potential from \citet{Usoskin-etal-2011} and not \citet{Usoskin-etal-2017} because we need a potential which reflects the solar modulation at neutron monitor rigidities. In \citet{Usoskin-etal-2017} the neutron monitor response has been calibrated to the direct PAMELA proton measurements thus reflecting these rigidities (cf. Sect.~\ref{section:solar-modulation-potential-for-different-rigidity-ranges} and especially Fig.~\ref{fig:phi_comparison_2006_2010_bu_vos}). 
In Fig.~\ref{fig:pam_phis_wrt_rigidity_bu_detail2} we already showed that the rigidity dependence of $\phi$ can be approximated by a straight line (in linear scale). Accordingly, we assume a linear interpolation between our two parameters, $\phi_{pp}$ and $\phi_{Uso11}$. 
Figure~\ref{fig:pam_phis_wrt_rigidity_bu_detail2} also shows the measured $\phi_{pp}$ and $\phi_{Uso11}$ for the first, mid and last PAMELA time period including their uncertainties as given in Sect.~\ref{section:solar-modulation-potential-for-different-rigidity-ranges} (red data points). 
Each pair of $\phi_{pp}$ and $\phi_{Uso11}$ is then connected by a straight line (in linear space), yielding our new rigidity-dependent modulation parameter $\phi(P)$ (magenta line):
\begin{align}
\phi(P) = 
\begin{cases} 
\frac{\phi_{Uso11}-\phi_{pp}}{P_{Uso11}-P_{pp}}\cdot(P-P_{pp})+\phi_{pp} & \text{if } P < P_{Uso11} \\
\phi_{Uso11} & \text{if } P \geq P_{Uso11}
\end{cases}
\label{eq:rig-dep-phi}
\end{align}
with $P_{pp}=1.28\pm0.01$~GV and $P_{Uso11}=13.83\pm4.39$~GV as derived in Sect.~\ref{section:proton-proxies} and \ref{section:rigidity-dependence-of-the-solar-modulation-potential}, respectively. Note that $\phi(P) = \phi_{Uso11}$ for higher (i.~e. neutron monitor) rigidities, where the minimization process to derive $\phi$ is not reliable (see Sect.~\ref{section:rigidity-dependence-of-the-solar-modulation-potential}).
The lower limit of $\phi(P)$ is given by the validity of the force field approach at lower rigidities. As described in Sect.~\ref{section:Cosmic ray transport in the heliosphere}, at 1~AU the force field solution starts to show significant deviations from a full numerical solution at energies below approximately 150-550~MeV, i.~e. 0.55-1~GV. Because of that the two parameter approach presented here - like the standard force field approximation - should not be applied below these rigidities.

Figure~\ref{fig:pam_phis_wrt_rigidity_ratios} shows the deviation of the rigidity-independent $\phi_{Uso11}$ (top) and $\phi_{pp}$ (middle), as well as the rigidity-dependent $\phi(P)$ (bottom) to $\phi_{PAM}$. Here $\phi_{PAM}$ is the solar modulation potential derived from PAMELA proton measurements for each rigidity bin and monthly measurement interval from Fig.~\ref{fig:pam_phis_wrt_rigidity_bu_detail2}. 
Figure~\ref{fig:pam_phis_wrt_rigidity_ratios} illustrates the rigidity ranges for which $\phi_{Uso11}$ and $\phi_{pp}$ are valid (i.~e. the deviation vanishes): $\phi_{Uso11}$, derived from neutron monitor measurements, is only able to reproduce $\phi_{PAM}$ measured by PAMELA at rigidities above 10~GV. And $\phi_{pp}$ can only be used up to 4~GV without deviating significantly from $\phi_{PAM}$. 
However, our newly derived rigidity-dependent modulation parameter $\phi(P)$ is able to describe the $\phi_{PAM}$-values obtained from PAMELA proton spectrum measurements over the whole rigidity range from 0.4~GV up to neutron monitor rigidities at 15~GV, where the minimization process becomes unreliable with increasing rigidities. 

\subsection{Comparison with other measurements}
\label{section:comparison_measurements}
\begin{figure*}
\includegraphics[width=\linewidth]{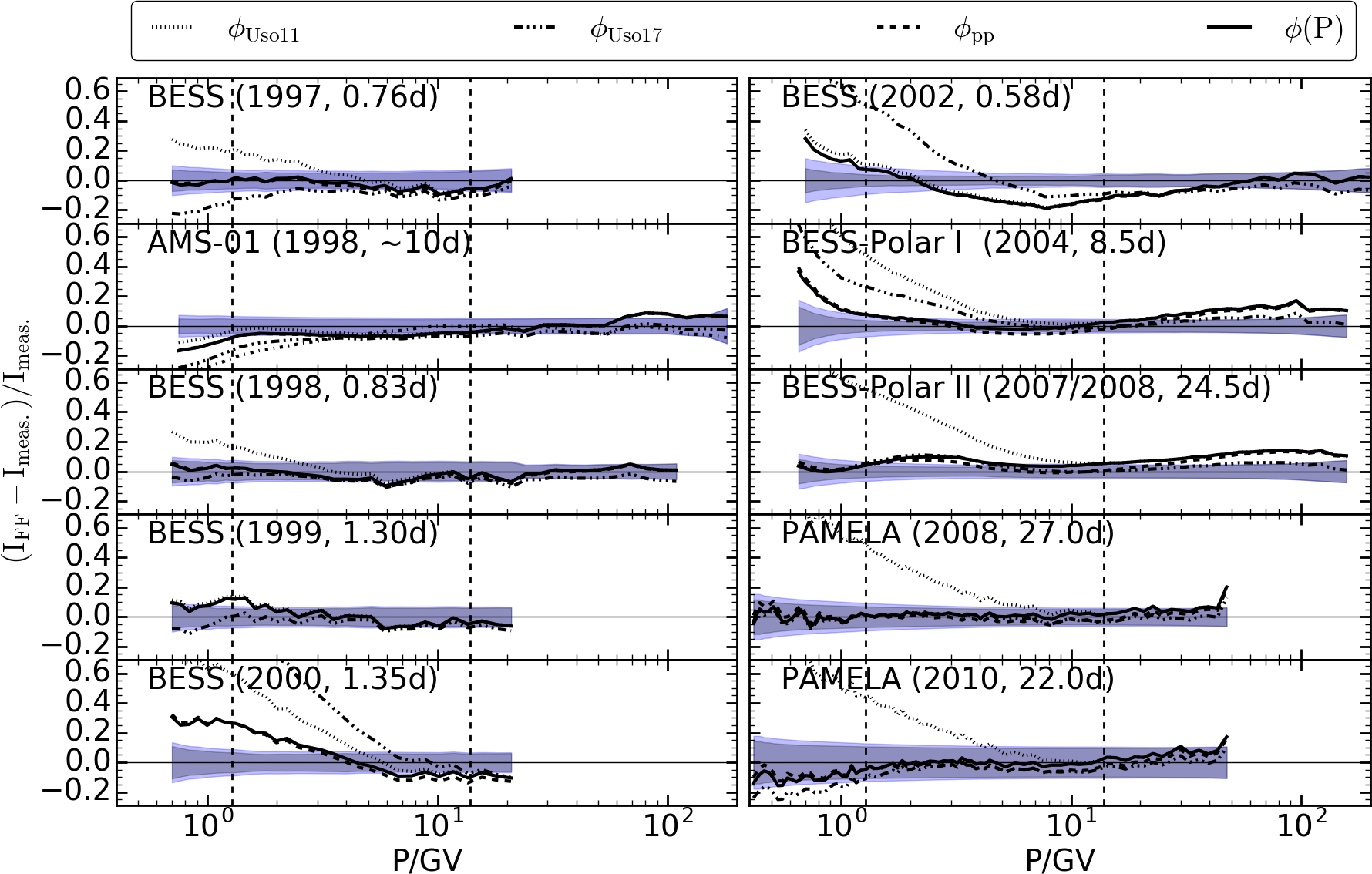}
\caption{
Deviations of proton spectra measured by AMS-01, BESS, BESS-Polar and PAMELA to force field models for solar modulation potentials from
\citet{Usoskin-etal-2011} ($\phi_{Uso11}$, dotted lines), 
\citet{Usoskin-etal-2017} ($\phi_{Uso17}$, dashed-dotted lines),
derived from proton proxies ($\phi_{pp}$, dashed lines), and 
from Eq.~\ref{eq:rig-dep-phi} ($\phi(P)$, solid lines), respectively. The shaded areas indicate the uncertainties for the zero deviation lines, resulting from the statistical and systematic errors of the measured intensities (inner shading) and the uncertainties of the $\phi$ calculations (additional outer shadings). 
Note that the time intervals of measurements (given in days next to the year) and models (i.~e. $\phi$-values, monthly values) can differ significantly, especially for the BESS missions. 
The measurement periods are also indicated in time series plot Fig.~\ref{fig:jgr_icrc2015_pamuso}.
}
\label{fig:verification}
\end{figure*}
To test our two parameter force field approach, a comparison with independent measurements of proton spectra at different times is given in Fig.~\ref{fig:verification} where the deviation between model and measurement is displayed. The model intensities are calculated using Eq.~\ref{eq:force-field-equation} with the LIS from \citet{Burger-etal-2000} and $\phi(P)$ from Eq.~\ref{eq:rig-dep-phi}, taking the monthly solar modulation potentials $\phi_{pp}$ and $\phi_{Uso11}$ as shown in Fig.~\ref{fig:phi_comparison_1973_2017} (and available as data set S1 in the supporting information).  
The measurements are from AMS-01 \citep{Alcaraz-etal-2000}, BESS \citep{Shikaze-etal-2007} and BESS-Polar I+II \citep{Abe-etal-2016}, partly obtained from the Database of Charged Cosmic Rays \citep{Maurin-etal-2014}.
In addition, two PAMELA measurements, which were used in our previous analysis to determine $\phi_{pp}$ and $\phi(P)$, are also included here.  
Note that the time intervals of measurements (given in days next to the year) and models (i.~e. $\phi$-values) can differ significantly, especially for the BESS missions. The BESS balloon flights each took measurements over roughly a day, whereas the solar modulation potentials were calculated for the whole corresponding month. This can result in significant deviations. For BESS-Polar II a weighted mean of the $\phi$-values for the mission time (December 2007 and January 2008) has been calculated. The shaded areas 
indicate the uncertainties for the zero deviation lines, resulting from the statistical and systematic errors of the measured intensities (inner shading) and the uncertainties of the $\phi$ calculations (additional outer shadings). 
For an interpretation of Fig.~\ref{fig:verification}, it is important to also take a look at the long-term temporal variation of GCRs, presented in Fig.~\ref{fig:jgr_icrc2015_pamuso} (top). Here the proton proxy time series is plotted, and the time intervals of all measurements shown in Fig.~\ref{fig:verification} as well as the solar polarity cycles are indicated. 
In the following each panel of Fig.~\ref{fig:verification} is interpreted in detail:
\begin{enumerate}
\item The three measurements by BESS and AMS from 1997 and 1998 were all obtained during an A$>$0 solar cycle and show similar results: for the BESS measurements, the $\phi_{Uso11}$-model delivers too high, the $\phi_{Uso17}$-model too low intensities at lower investigated rigidities, whereas the $\phi(P)$-model shows the lowest deviation at these rigidities; for AMS 1998 all models yield slightly too low intensities and are quite close to each other.
\item Measurements from BESS 1999 were made in the declining phase of the A$>$0 epoch and show similar results for all models, with the $\phi_{Uso17}$-model giving slightly lower deviations.
\item The BESS 2000 and 2002 observations provide less good agreements between models and measurement because they took place during the solar maximum. However, there are big differences between the models at lower rigidities. The $\phi(P)$-model yields deviations of up to 30\% in 2000 and 20\% in 2002, while the $\phi_{Uso11}$-model deviates by up to 60\% and 20\%, and the $\phi_{Uso17}$-model by more than 60\% for the same periods.
\item BESS-Polar I was launched 2004 in the rising phase of the last A$<$0 solar cycle. The deviations show similar behavior as before: at higher rigidities all models show comparable and good results, while below 5~GV only the $\phi(P)$-model gives deviations of less than 10\%. As in 2002, for the very low investigated rigidities all models provide too high intensities.
\item The comparisons in 2007, 2008, and 2010 with BESS-Polar II and PAMELA, respectively, all took place during the last A$<$0 epoch and yield comparable results: the models for $\phi(P)$ and $\phi_{Uso17}$ show good agreements with the measurements, while the $\phi_{Uso11}$-model shows big deviations below 5~GV.
\end{enumerate}
In summary, we state that for all analyzed time intervals from 1997 to 2010, covering the last A$>$0 epoch and its declining phase as well as the last A$<$0 epoch with its rising phase, the $\phi(P)$-model yields the lowest deviations and delivers for almost all cases good results. 
The $\phi_{Uso17}$-model shows good agreements with the measurements for all observations in the A$<$0 solar cycle, at which it has been calibrated to the measurements. But it has the biggest deviations in the solar maximum phase and underestimates the intensities in the last A$>$0 solar cycle. 
The $\phi_{Uso11}$-model can only describe the measurements at all rigidities during some intervals in the last A$>$0 solar cycle and for one measurement at solar maximum; apart from that it shows big deviations below 5~GV.

\section{Importance of the new solar modulation potential values for the production rate values of $^{10}$Be}
\label{section:cosmogenic-isotopes-production-rates}
 \begin{figure}[t!]
 \includegraphics[width=\columnwidth]{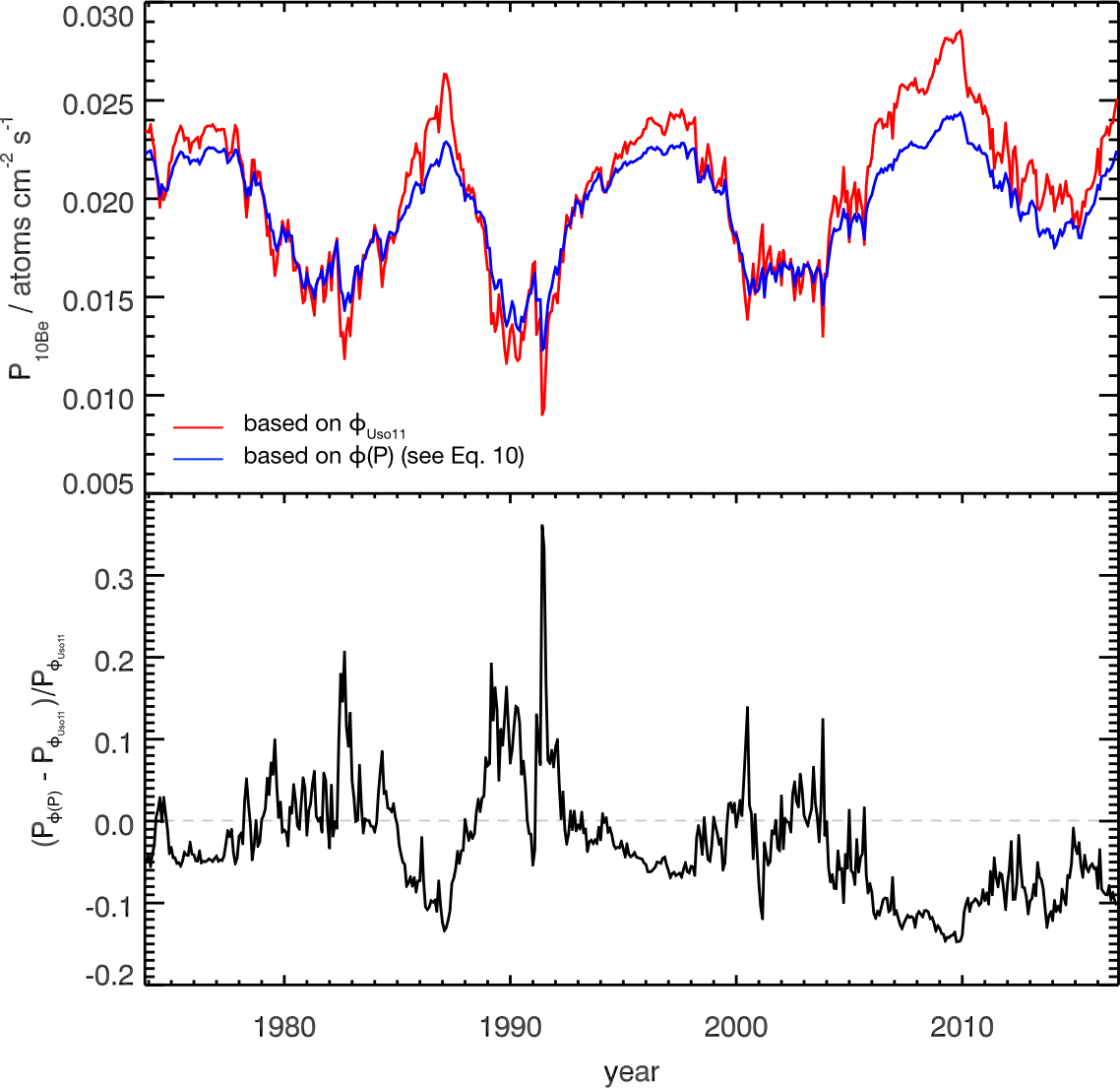}
 \caption{ Top: Time profiles of global $^{10}$Be production rate values based on the solar modulation parameter by \cite{Usoskin-etal-2011} (red line) and $\phi(P)$ from Eq.~\ref{eq:rig-dep-phi} (blue line). Bottom: Deviation of these two production rate values.}
 \label{fig:10be}
 \end{figure}
The production of secondary particles within the Earth's atmosphere strongly depends on the GCR flux interacting with the atmospheric constituents. These secondary particles may also produce so-called cosmogenic radionuclides like $^{10}$Be, $^{14}$C and $^{36}$Cl, which are often used as a proxy of the solar activity on time scales of thousands of years \citep[see e.~g.][]{Steinhilber-etal-2012, Muscheler-etal-2016}. Since there is an anti-correlation between the production and the solar activity, the force field solution and its solar modulation potential $\phi$ is commonly used to compute the production rate values. For further information on the computations itself see e.~g. \citet{Herbst-etal-2017}.

In order to estimate the influence of changes in the solar modulation potential due to the analysis of its rigidity dependence as described in Sect.~\ref{section:rigidity-dependence-of-the-solar-modulation-potential}, we compute the global production rate values of $^{10}$Be for the newly presented rigidity-dependent $\phi(P)$ as defined by Eq.~\ref{eq:rig-dep-phi}. The upper panel of Fig.~\ref{fig:10be} shows the temporal evolution of the global $^{10}$Be production based on the solar modulation potential by \cite{Usoskin-etal-2011} (red line) as well as the newly derived $\phi(P)$ (blue line). As discussed in \cite{Herbst-etal-2017}, a direct comparison is only possible because both records are based on the same LIS model \citep[here][]{Burger-etal-2000}. It shows that the results displayed in Fig.~\ref{fig:phi_comparison_1973_2017} are reflected in these computations. While the production rate values are in good agreement during times of similar solar modulation potential values (i.~e. around 1980 and 2000) strong deviations occur when the $\phi$-values strongly differ from each other (i.~e. around 1990 and 2010). The bottom panel of Fig.~\ref{fig:10be} shows the deviation between  the production rates based on $\phi(P)$ and $\phi_{Uso11}$, respectively. Non-negligible differences of more than $\pm15\%$ occur. This is of great importance, because the solar activity reconstructions from the cosmogenic radionuclide records, which go back thousands of years, are based on the solar modulation values during the spacecraft era. Utilizing the newly reconstructed rigidity-dependent $\phi(P)$-values may have a strong influence on these reconstructions.

Note, that this analysis has not been performed for the newer solar modulation potential values from \citet{Usoskin-etal-2017} based on the LIS by \citet{Vos-Potgieter-2015}. As noted before due to its derivation $\phi_{Uso17}$ reflects the solar modulation as observed by PAMELA proton measurements and not by neutron monitors. Thus, using $\phi_{Uso17}$ we lack information of the solar modulation at higher rigidities and, therefore, we cannot derive a rigidity-dependence based on this parameter.

\section{Summary}
\label{section:summary}
In this work we have demonstrated that the commonly used force field approach shows a significant rigidity dependence below 10~GV. As a simple yet sufficient workaround we introduced a modification to the model using two solar modulation potential parameters determined by GCR measurements at different rigidities. Thus, we were able to provide a monthly rigidity-dependent solar modulation potential for the period from 1973 to 2016, covering two A$>$0 and two A$<$0 solar magnetic cycles. 
It can easily be calculated following Eq.~\ref{eq:rig-dep-phi} using the data set S1 included in the supporting information of this manuscript (also online at \url{http://www.ieap.uni-kiel.de/et/ag-heber/cosmicrays}).

In order to obtain the solar modulation potential for protons at around 1.28~GV for the whole time period of more than 40 years, we took advantage of the fact that different GCR ions show the same temporal behavior in the inner heliosphere if compared at the same rigidity. Thus, we could use IMP-8 helium and ACE/CRIS carbon measurements, normalized to PAMELA proton observations, to obtain so called 1.28~GV proton proxies, which were then used to calculate the corresponding solar modulation potential at these lower rigidities. In addition, we utilized the solar modulation potential calculated by \citet{Usoskin-etal-2011} from neutron monitor measurements for higher rigidities. Compared to newer findings by the same authors \citep{Usoskin-etal-2017}, \replaced{these}{this} potential was found to reflect the neutron monitors observations best.
These two modulation potentials at mean rigidities of approximately 1.28~GV and 13.83~GV already demonstrate the rigidity dependence of the force field approach. 
The significance of this dependency has been emphasized by the full rigidity-dependent solar modulation function, which we calculated from PAMELA proton spectra from 2006-2010 by 
connecting our two modulation potentials from spacecraft and neutron monitor observations with an empirical rigidity transition function.
We compared the rigidity-independent modulation potentials from \citet{Usoskin-etal-2011}, \citet{Usoskin-etal-2017} and from our proton proxies together with the newly derived rigidity-dependent potential function to independent observations by AMS-01, BESS and BESS-Polar (as well as dependent PAMELA measurements) for time periods from 1997 to 2010. Thereby, we could demonstrate that the here presented force field modification is the only model in the comparison which is able to describe the observed proton spectra from 1-100~GV in both solar polarity cycles. 

The impact of the different solar modulation potentials on the production rates of the cosmogenic radionuclide $^{10}$Be has been illustrated at the end of this work. We showed that the production rate values based on our newly developed rigidity-dependent solar modulation potential $\phi(P)$ have a non-negligible difference of more than $\pm15\%$ from those based on the record by \cite{Usoskin-etal-2011}, especially during  solar minima. And although our simplified model cannot replace full numerical solutions of the transport equation, which are necessary to further investigate all propagation processes of GCRs in the heliosphere, it is an easy to use (and within its limitations reasonable) two-parameter model to describe the solar modulation during the spacecraft era, on which all solar modulation reconstructions from cosmogenic radionuclide records are based on.

Upcoming updated data sets from PAMELA and AMS-02 (with potentially higher precision) for time periods after 2010 will help to investigate the rigidity-dependence of the solar modulation potential in more detail. In addition, they will improve the normalization of the proton proxies by covering not only a part of but the full A$<$0 and also the beginning of the next A$>$0 solar cycle.

\acknowledgments
This work was partly carried out within the framework of the bilateral BMBF-NRF-project "Astrohel" (01DG15009) funded by the Bundesministerium f\"ur Bildung und Forschung. 
We thank the PAMELA collaboration for providing the proton data via the Italian Space Agency (ASI) Science Data Center (\url{http://tools.asdc.asi.it/cosmicRays.jsp}).
We thank the ACE/CRIS instrument team and the ACE Science Center for providing the ACE data (\url{http://www.srl.caltech.edu/ACE/ASC/}).
We acknowledge the NMDB database (\url{http://www.nmdb.eu}), founded under the European Union's FP7 program (contract no. 213007) for providing the Kiel neutron monitor data.
\added{We thank the Sodankyl\"a Geophysical Observatory of the University of Oulu for providing the solar modulation potential from \citet{Usoskin-etal-2005,Usoskin-etal-2011} (\url{http://cosmicrays.oulu.fi}).}
Sunspot number data used in this study was obtained via the web site \url{http://www.sidc.be/silso/} courtesy of the SIDC-team, World Data Center for the production, preservation and dissemination of the international sunspot number, Royal Observatory of Belgium.
We thank the Database of Charged Cosmic Rays (\url{http://lpsc.in2p3.fr/crdb}) for providing easy access to multiple cosmic ray data sets.
J.~G. would like to thank M. Potgieter and D. Strauss for many fruitful discussions. 
The data presented in Fig.~\ref{fig:phi_comparison_1973_2017} is included in the supporting information for this manuscript (also online at \url{http://www.ieap.uni-kiel.de/et/ag-heber/cosmicrays}).
\def\aj{AJ}%
\def\actaa{Acta Astron.}%
\def\araa{ARA\&A}%
\def\apj{ApJ}%
\def\apjl{ApJ}%
\def\apjs{ApJS}%
\def\ao{Appl.~Opt.}%
\def\apss{Ap\&SS}%
\def\aap{A\&A}%
\def\aapr{A\&A~Rev.}%
\def\aaps{A\&AS}%
\def\azh{AZh}%
\def\baas{BAAS}%
\def\bac{Bull. astr. Inst. Czechosl.}%
\def\caa{Chinese Astron. Astrophys.}%
\def\cjaa{Chinese J. Astron. Astrophys.}%
\def\icarus{Icarus}%
\def\jcap{J. Cosmology Astropart. Phys.}%
\def\jrasc{JRASC}%
\def\mnras{MNRAS}%
\def\memras{MmRAS}%
\def\na{New A}%
\def\nar{New A Rev.}%
\def\pasa{PASA}%
\def\pra{Phys.~Rev.~A}%
\def\prb{Phys.~Rev.~B}%
\def\prc{Phys.~Rev.~C}%
\def\prd{Phys.~Rev.~D}%
\def\pre{Phys.~Rev.~E}%
\def\prl{Phys.~Rev.~Lett.}%
\def\pasp{PASP}%
\def\pasj{PASJ}%
\def\qjras{QJRAS}%
\def\rmxaa{Rev. Mexicana Astron. Astrofis.}%
\def\skytel{S\&T}%
\def\solphys{Sol.~Phys.}%
\def\sovast{Soviet~Ast.}%
\def\ssr{Space~Sci.~Rev.}%
\def\zap{ZAp}%
\def\nat{Nature}%
\def\iaucirc{IAU~Circ.}%
\def\aplett{Astrophys.~Lett.}%
\def\apspr{Astrophys.~Space~Phys.~Res.}%
\def\bain{Bull.~Astron.~Inst.~Netherlands}%
\def\fcp{Fund.~Cosmic~Phys.}%
\def\gca{Geochim.~Cosmochim.~Acta}%
\def\grl{Geophys.~Res.~Lett.}%
\def\jcp{J.~Chem.~Phys.}%
\def\jgr{J.~Geophys.~Res.}%
\def\jqsrt{J.~Quant.~Spec.~Radiat.~Transf.}%
\def\memsai{Mem.~Soc.~Astron.~Italiana}%
\def\nphysa{Nucl.~Phys.~A}%
\def\physrep{Phys.~Rep.}%
\def\physscr{Phys.~Scr}%
\def\planss{Planet.~Space~Sci.}%
\def\procspie{Proc.~SPIE}%
\listofchanges

\end{document}